\documentclass[conference]{IEEEtran}
\IEEEoverridecommandlockouts
\usepackage{graphicx}
\usepackage{epstopdf}
\usepackage{placeins}
\usepackage{url}
\usepackage{amssymb,amsmath}
\usepackage{tabularx}
\usepackage{array}
\usepackage{todonotes}
\usepackage{verbatim}
\usepackage{todonotes}
\usepackage{multirow}
\usepackage{hhline}
\usepackage{balance}

\usepackage{paralist}
\usepackage{enumerate} 
\usepackage{epstopdf}

\newcommand{\feedback}[1]{\textbf{#1}}
\newcommand{\ncn}[1]{\feedback{NCN: #1}}

\begin{document}

\title{Managing Large-Scale Transient Data in IoT Systems}
%Provisioning and Managing Resources in IoT Systems: The Information-Centric Way(\linh{maybe: An Information-Centric Approach for Provisioning an End-to-End Network of Mobile-Edge Cloud Resources})
%}
\author{
   \IEEEauthorblockN{Nanjangud C.~Narendra}
    \IEEEauthorblockA{Ericsson Research Bangalore, Bangalore, India\\
      nanjangud.narendra@ericsson.com}
    \and
   \IEEEauthorblockN{Sambit Nayak}
   \IEEEauthorblockA{
    Ericsson Research, Bangalore, India\\
     sambit.nayak@ericsson.com }
		\and
		\IEEEauthorblockN{Anshu Shukla}
   \IEEEauthorblockA{
    Ericsson Research, Bangalore, India\\
     anshu.shukla@ericsson.com}
}

\maketitle

\begin{abstract}

The rise of Internet of Things (IoT) has brought about the need to manage the voluminous amount of data that flows through IoT systems. In order to achieve scalability, distributed cloud technology is used for designing and implementing large-scale IoT systems. Current work on managing data in such systems has mostly focused on persistent data, i.e., data that is stored even after the system has finished its execution. However, very little work has focused on how to manage the transient (non-persistent) data that is streamed through the system and which does not outlive the system execution. This transient data is crucial since it is typically processed to generate summarized insights during system execution, and some of it may need to be stored after system execution as per users’ needs. Currently this data is either purged after analysis or is fully stored for historical purposes.

To that end, in this position paper, we present our thoughts on managing transient data in IoT systems. By managing, we mean the process of generating useful value from this data for users. We therefore provide a precise definition of how data can be characterized as transient. We then use this definition to suggest approaches for facilitating placement and processing of this data closer to related data and/or sources of computation for improving execution efficiency. We also show how these approaches can be made dynamic, so as to accommodate users' changing needs in real-time.

\end{abstract}

\section{Introduction and Research Challenges}\label{sec:intro}

The world is witnessing an exponential growth in the number of devices comprising sensors, actuators and data processors, resulting in the so-called Internet of Things (IoT) phenomenon. It is estimated that by 2020 there will be in excess of 50 billion devices all connected to the Internet. Such a proliferation of devices has the potential to generate enormous amounts of data, resulting in a classic Big Data problem, viz., the need to extract, process and analyze this data. In addition, most of this data is \emph{transient}; in other words, it is usually generated at high speeds by sensors and other devices, and its usefulness reduces rather rapidly with time.

%Consider, for example, temperature sensors on a floor of an office building. They would typically generate temperature readings every few minutes depending on how they are programmed. This data typically goes through gateways and edge cloud servers, through which it is propagated, and the data is then sent to a central cloud server for analysis. Typical analysis of this data would include calculating average, maximum and/or minimum temperatures on the floor. More detailed analysis could involve determining temperature profiles, such as distribution of temperature across the floor, and correlation thereof with sunlight streaming into the building. However, in current IoT systems, most of these analyses are conducted in backend cloud servers after all the data is propagated there. While the results of these analyses are probably stored depending on users' specific requirements, practically all of the transient data is usually discarded. 

Consider, for example, an offshore oil platform (see https://www.automationworld.com/oil-and-gas-edge). The sensors within it typically generate 1-2 TB of data. Given its relatively remote geographic location, it would take dedicated satellite links to upload that data onto the cloud, which is infeasible given the relatively low speed of satellite uploads. It is estimated it would take several days just to upload one day of data. Since all this data needs to be analyzed, it is essential that the refinery be equipped with appropriate tools close to the edge so that most of the data can be analyzed, and decisions taken about which data to retain \& transmit, in real-time. In addition, latency is also an issue, since data that is not used immediately gets lost, and that degrades the quality of the overall analysis conducted on the data. Such a scenario calls for a locally designed and installed IoT network within the oil platform, that is able to quickly process this data and send only a subset of the data to the remote cloud for storage \& further processing. 

Another interesting example is in transportation systems in smart cities~\cite{Giang2016}, in particular, Vehicular Ad-Hoc Networks (VANET). Such networks generate huge amounts of data (several GB per mile) which needs a well-designed network of intermediate edge servers and data aggregators in order to be able to process and derive value from this data. Also, unlike in the case of the oil platform, such networks would need to be dynamically modified in response to changing vehicular traffic and other road conditions. Some applications here are traffic light management and vehicular search (searching for a suspect driving a vehicle with the help of dashboard cameras mounted on nearby vehicles).

These examples therefore raise a number of research questions that need to be addressed, of which we list a few:

\begin{enumerate} 
\item How to define which data is transient, i.e., which data needs to be discarded after analysis?
\item How to design the IoT network and its intermediate transient data storage and analysis servers on the edge so that the expected volume of data can be handled?
\item How to optimally place transient data throughout the IoT network so that the overall latency of data transmission and analysis is minimized?
\item How to ensure that the data placement strategy is dynamic; as a corollary, how to minimize the time taken to migrate data in case of this dynamism?
\item How to determine, on the fly, that subset of data in an intermediate data server, that needs to be sent to a central cloud server for storage? This could depend on its relation to other data sources in other servers throughout the IoT network. 
\item How to use provenance to determine which data to be sent to the central cloud server? Provenance, i.e., history of past usage, can, in addition to the actual nature of the data itself, dictate which data needs to be stored in the central cloud server.
\end{enumerate}

\section{Our Proposed Approach}\label{sec:approach}

\begin{comment}

\ncn{Some aspects of our proposed approach:
\begin{itemize}
\item overall view of the IoT network in terms of data sources and data sinks
\item knowledge of the data transmitted by the data sources, comprising: data type, unit, frequency of transmission
\item initial layout of overall IoT network, i.e., topology of data sources and sinks, in order to achieve minimal latency - need to cite a few papers from Yogesh here as well as some of the papers that we uncovered in our own research \& which are used in our various MSc proposals
\item DSL specification by the user, which specifies the following: for each data item, first classify if transient; if so, how much of it needs to be stored
\item Based on the above specification, the system will calculate - via a Resource Estimation Engine - storage requirements of the transient data sets involved in a data pipeline/flow and the compute requirements of the involved processing units
\item Furthermore, the Resource Planning Engine will perform matching of the estimated storage and compute resource requirements with the available capabilities at the various distributed locations
\item In parallel, the Data Management Engine will perform background actions of deducing associations between data sets, tracking data provenance, and deriving possible optimizations such as temporary data migration or transformation for a data set
\end{itemize}
}

\end{comment}

\begin{figure*}[htbp]
\centering
\includegraphics[width=7in,height=5.5in]{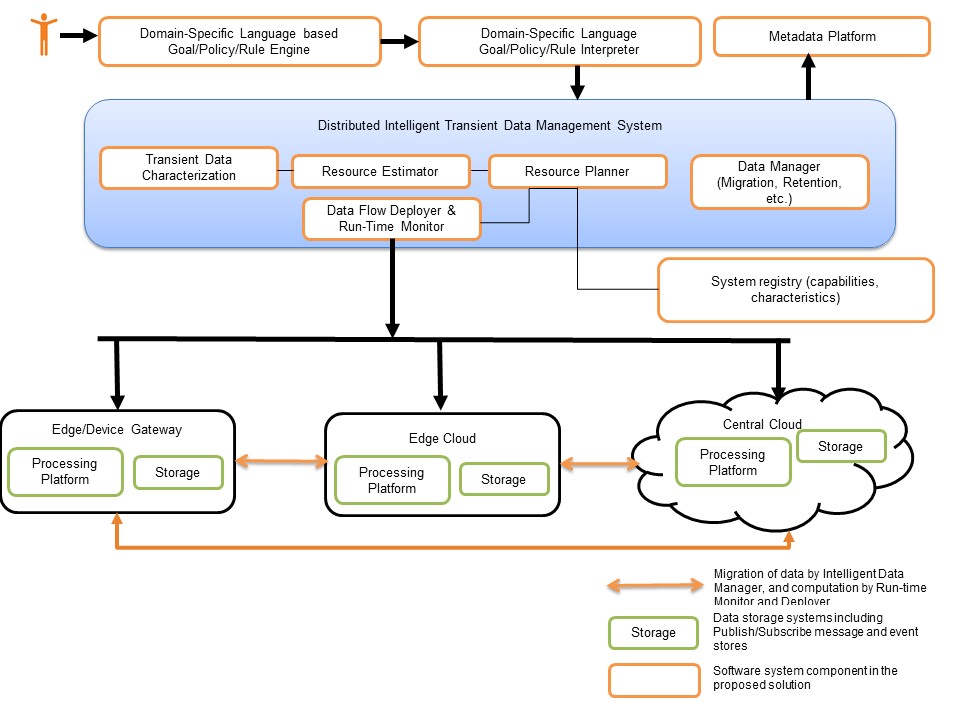}
% where an .eps filename suffix will be assumed under latex,
% and a .pdf suffix will be assumed for pdflatex; or what has been declared
% via \DeclareGraphicsExtensions.
\caption{System Architecture}\label{fig:sysarch}
\end{figure*}

Our proposed system for transient data management is as depicted in Fig.~\ref{fig:sysarch}, and comprises the following components:
\begin{itemize}
\item \emph{Transient Data Characterization} is responsible for deducing which data sets can be categorized as “transient” (though not explicitly marked so by the user) and their associated storage characteristics
\item \emph{Resource Estimator} is responsible for estimating storage requirements of the transient data sets involved in a data pipeline/flow and the compute requirements of the involved processing units
\item \emph{Resource Planner} performs matching of the estimated storage and compute resource requirements with the available capabilities at the various distributed locations
\item \emph{Data Manager} performs background actions of deducing associations between data sets, tracking data provenance, and deriving possible optimizations such as temporary data migration or transformation for a data set
\end{itemize}

%We now need to detail out the above along with the diagrams, 

%Fig.~\ref{fig:sys-op} shows the overall dynamic behavior of our system:
\begin{itemize}
\item The system maintains a registry of available storage and compute capabilities at the edge (constrained or more-capable), fog and cloud systems/nodes.
\item The user can author rules or policies in a Domain-Specific Language (DSL) provided by the system to describe a high-level data processing and flow goal. The system parses such higher-level goals to essentially derive a description of the data sets involved and any processing and flow pipeline elements. The user can also directly categorize certain data sets as ``transient'' in the system, or the system can deduce which data sets are transient from the goals themselves. Characterization of transient data involves identifying its characteristics like generation source(s), validity period, estimated storage volume, etc.
\item Once such characterization of transient data is done, the system performs matching of available storage capabilities at the known/registered systems/nodes with the required estimated capability for the transient data sets, and thus allocates such storage as close to the data generation source as possible.
\item The system performs similar estimation and matching for the compute requirements of the data processing/flow pipeline units.
\item The system registers metadata information about the transient data sets into the metadata platform.
\item The system formulates the data flow as a Direct Acyclic Graph (DAG) model comprising of computation and storage units, based on the identified data processing elements and data sets involved. And then the system deploys the DAG units as per the model.
\item The system runs background actions to:
	\begin{itemize}
	\item monitor the DAG units and the systems/locations they are deployed on to perform run-time estimation of any changes in required storage and compute resources, match them dynamically with available storage and compute, and do necessary re-deployment of the updated DAG
	\item analyze provenance of and associations between data sets, and their historical access patterns and thereby perform data management actions (migration, transformation, retention, etc.) as necessary
	\end{itemize}
\end{itemize}

%\begin{figure}[htbp]
%\centering
%\includegraphics[width=3.5in,height=2.5in]{figures/sys-op.jpg}
% where an .eps filename suffix will be assumed under latex,
% and a .pdf suffix will be assumed for pdflatex; or what has been declared
% via \DeclareGraphicsExtensions.
%\caption{System Operation}\label{fig:sys-op}
%\end{figure}

We now describe the primary facets of the overall system described above, with the individual facets and their features described in detail with illustrative examples:
\begin{itemize}
\item Characterization as Transient Data: ``Transient data'' refers to data that is not primarily intended for long-duration storage, rather certain kinds of processing/transformations are done on such data to derive resultant data that is useful to the user or system, and the resultant data in turn is stored or archived as per requirements. Such categorization of data as ``transient'' can be performed based on direct user actions and/or policies set by the user. Alternatively, the system itself deduces that certain types of data are ``transient'' in nature - either learning from past record of explicit marking or on its own. An example would be temperature readings in a building, of which only summarized data would be subject to analysis by earlier users.
\item Provenance-based intelligent data management: Use of a ``metadata'' platform~\cite{hellerstein2017ground} by the system will facilitate intelligent decisions and platform-level data management actions on such data by the system. Below we describe certain kinds of intelligent actions. For example, the system may maintain associations between data sets to build knowledge of related data; or, it can store and process transient data as close to the point of generation as possible; or it can analyze historical data set access patterns to perform any possible data management actions in the background for optimization purposes.

\noindent Such placement of data and computation is also dynamically adjusted at run-time based on the available capabilities (which can change over time with or without configuration changes), policies \& quality parameters (configuration changes by user). That means the system performs on-the-fly migration of transient data sets and computation logic among the possible placement locations. Such migration is based on a set of decision criteria – such as data transfer and storage costs, user-defined policies/rules, quality parameters such as overall latency and capacity of Virtual Machines, etc.
\item Rule-based Domain Specific Language (DSL) for Configuration: Policies/rules authored by the user will be in a high-level domain-specific language (DSL) that states the goals, rather than explicitly stating the technical specifics of data storage and processing. The system will deduce the optimal storage and computation requirements for a data flow described in such a goal, and match them with available capabilities while attempting to optimize costs. For example, a higher level goal in DSL would state that ``collect temperature data throughout a building every 30 seconds, send a notification to all hooters/signals if temperature crosses a threshold T, and store the resultant summarized metrics (average over every 15 minutes) and alarms data only to a target system every 24 hours for further analysis''. In this case, the system will: identify that the sensor readings accumulated over a 30-second period is "`transient data'' and that the averages and alarms/events data is also ``transient''; perform estimation of the volume of transient data over the 15-minute window, and match that with the available storage/compute capability at the relevant edge gateway or edge/cloud server; and perform estimation of the computation requirements (pull/fetch from sensor or pub/sub system, summary/average and event/alarm generation, write to storage, push to target system) and match such requirements with computation capabilities at the edge gateway and/or edge/cloud server closest to the temperature sensors/devices in the building deployment; based on these estimations, deplys the data flow processing units accordingly.

\end{itemize}

\section{Related Work}\label{sec:relatedwork}

\noindent\emph{Data Modeling}: The citation~\cite{George2009} presents spatio-temporal sensor graphs, a data model for representing sensor data. Its key usefulness is a memory-efficient model for representing fast-changing sensor data, that also supports adequate support for knowledge discovery from the model. The model also enables the detection of so-called ``hot spots'', which are topological changes occurring in the graph, and are generated by changes in values emitted by sensors. The spatio-temporal sensor graph is therefore a useful tool for representing changes in sensor values over time, and is complementary to our work.

\noindent\emph{Data Processing}: The citation~\cite{Cecchinel2014} presents an architecture for a data collection system for IoT based applications. That paper also presents an implementation and detailed experimental results to evaluate their architecture. A similar data collection \& storage architecture, specific to smart cities, is presented in~\cite{Fazio2014}. It provides a high abstraction layer for the description of both sensing infrastructures and sensed data, which can be exported as ``things'' to the internet.

The CityPulse project (see http://www.ict-citypulse.eu/) focuses on providing large-scale stream processing solutions to interlink data from Internet of Things and relevant social networks and to extract real-time information for the sustainable and smart city applications. The project supports the integration of dynamic data sources and context-dependent on-demand adaptations of processing chains during run-time. We view CityPulse as complementary to our work, since our contextual data filtering approach would be relevant to any system that needs to process streaming data from sensors.

A recent trend in cloud computing has been Fog Computing~\cite{Fog2014} and Edge Computing (see http://vis.pnnl.gov/pdf/fliers/EdgeComputing.pdf). These proposals focus on the crucial issue of voluminous data generated by sensors, and which need to be processed by cloud data centers. Arguing that centralized cloud data centers would be overwhelmed by the volume of this data, these proposals call for moving computation and data analytics as much towards the edge devices as possible. 

\noindent\emph{Data Storage}: A context management architecture for IoT has also been presented in~\cite{Berbers2014}. One key feature of~\cite{Berbers2014} is a multi-tenant storage \& representation model that provides data isolation for multiple users. Another key feature is scalability, which is achieved via a combination of distributed deployment with horizontal scalability, and shared resources through multitenancy. We will be investigating the multitenancy approach of~\cite{Berbers2014} for our future work.

On a related note, the citation~\cite{Antunes2014} presents an approach for scalable semantic-aware storage of context data. The key premise of that paper is that, given the extreme heterogeneity of data sources in IoT-based systems, it would be more efficient to deal with this diversity via context-aware storage. To that end, that paper discusses the key requirements for context storage systems, and discusses two context organization models. Simulation of these models on smart city data is also presented in~\cite{Antunes2014}.

In an earlier paper~\cite{NarendraCCEM2015}, we presented a decentralized cloud-based solution for storage of IoT data, using Cloud and mini-Clouds. The key features of our paper are the proposal to use an object storage product such as Ceph;  algorithms for optimal mini-Cloud placement in an IoT environment; as well as algorithms for optimal migration of data among the mini-Clouds to address storage capacity issues while minimizing access latency. In~\cite{NarendraIEEECloud2016} we further explored this topic by investigating optimal data replication strategies in distributed IoT cloud storage. Our solution assumes the existence of multiple distributed cloud data centers, called mini-Clouds, among which data can be replicated. We model our problem comprehensively based on various parameters such as effective bandwidth of the IoT network, available number and size of data items at each mini-Cloud, and we present our problem as a collection of various sub-problems based on subsets of these parameters. We prove that the exact solution to the problem is intractable, and we present a number of heuristic strategies to solve it. Our results show that the performance of any heuristic is bounded by the read and write latency of mini-Clouds. We will be investigating incorporation of the techniques from~\cite{NarendraCCEM2015,NarendraIEEECloud2016} into our future work on transient data management.

\section{Conclusions and Future Work}\label{sec:conclusions}

In this paper, we have addressed the crucial issue of managing transient data in IoT systems. While much work has gone into data management in IoT in general, we show that a precise characterization of transience of data in such systems is currently lacking. To that end, we have provided in this paper a precise definition. Based on this definition, we proposed approaches by which this transient data can be placed \& processed for deriving useful value for users of this data. 

Future work will focus on implementing and testing our approaches on specifically designed IoT test beds in various application domains such as smart grid, building automation and transportation. 

%In this paper, we address the issues of managing heterogeneous and large-scale IoT resources. With the high-level view of IoT resources and the integration mechanism, our solution is extensible to incorporate the software-defined gateways and network functions. We shows that our framework can capture information from different data models from different providers. The distributed architecture and a management tool facilitate the resource management in both global and local level. The result of this work fosters further research such as IoT resource provisioning, dynamic configuration, and optimal distributed IoT data storage.

\section*{Acknowledgments} 
The authors wish to thank their colleagues at Ericsson Research Bangalore for their feedback.
%\balance
\bibliographystyle{IEEEtran}
\bibliography{refs}
\end{document}